# Ultrafast Spin-Resolved Spectroscopy Reveals Dominant Exciton Dynamics in Conducting Polymer Polyaniline

Soonyoung Cha,[†] Yoochan Hong,[‡] Jaemoon Yang,[§] Inhee Maeng,[†] Seung Jae Oh,[§] Kiyoung Jeong,[¦] Jin-Suck Suh,[§] Seungjoo Haam,[#] Yong-Min Huh,[*,§] and Hyunyong Choi[*,†]

[†] School of Electrical and Electronic Engineering, Yonsei University, Seoul 120-749, Republic of Korea

[‡] Department of Biomedical Engineering, Yonsei University, Wonju 220-740, Republic of Korea

[§] Department of Radiology, YUMS-KRIBB Medical Convergence Center, College of Medicine, Yonsei University, Seoul 120-752, Republic of Korea

[¦] Nanomedical National Core Research Center, Yonsei University, Seoul 120-749, Republic of Korea

[#] Department of Chemical and Biomolecular Engineering, Yonsei University, Seoul 120-749, Republic of Korea

**Abstract**

The conducting polymer polyaniline (PANI) has a wide range of optoelectronic applications due to its unique electronic and optical characteristics. Although extensive works have been performed to understand the equilibrium properties, the nature of the charge type that governs its non-equilibrium optical response has been barely understood; a number of studies have debated the nature of photo-generated charge type in PANI, specifically whether it is polaron or exciton based. Here, we report experimental evidence that the charge relaxation dynamics of PANI are dominated by excitons. Utilizing ultrafast spin-resolved pump-probe spectroscopy, we observed that PANI charge dynamics are strongly spin-polarized, exhibiting a spin Pauli-blocking effect. Investigations including both spin-independent and spin-dependent dynamics reveal that there is no spin-flip process involved in charge relaxation. This provides compelling evidence of an exciton-dominated photo-response in PANI.

**Table of Contents**

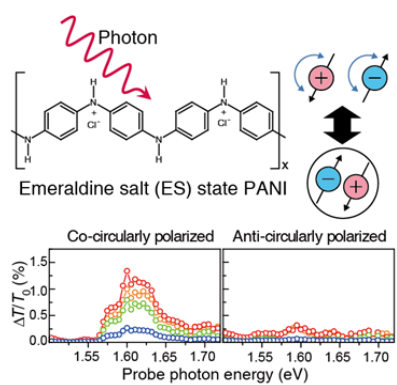

1. Introduction

Polyaniline (PANI), a conducting polymer, has received increasing attention from the chemistry and materials science communities over the last several decades due to its unique electronic and optical characteristics.[1-7] The metallic state of PANI, known as an emeraldine salt (ES) state, is a protonated form of its insulating counterpart, an emeraldine base (EB) state (Figure 1a).[1, 2] With insulator-to-metal transitions, electronic conductivity is significantly enhanced in the ES state PANI. More interestingly, the optical absorption of ES state PANI exhibits a strongly red-shifted feature compared to the EB state (Figure 1b,c).[4] Applications of such optical characteristics have led to developments such as polymer ink,[8] pH-sensitive detectors,[9] and electro-luminescence devices,[10] and so on.[11, 12] A recent optical study extended the applicability of PANI to biochemical applications, showing that its pH-sensitive optical properties can be employed for photo-thermal ablation of cancer cells.[13] While a large body of existing work has focused on its equilibrium optical characteristics, there has been little investigation of the dynamic photo-response of PANI and its carrier relaxation kinetics.

Although equilibrium optical spectroscopy has several benefits in characterizing the stationary properties of samples, ultrafast measurements on the photo-excited non-equilibrium dynamics provide direct time-resolved information in determining charge relaxation kinetics in polymers.[14] One important issue in polymer physics is whether the photo-induced charge excitation is dominated by polarons or excitons. This question is still actively debated despite extensive polymer research. This issue is important because charge-excitation type determines the range of photo-induced optoelectronic applications for conducting polymers. Prior studies[15, 16] using ultrafast pump and probe spectroscopy have shown that photon injection excites electrons from the valence band to the polaron band. The polaron state is an intermediately doped state lying below the conduction band edge (Figure 2a). These studies have suggested polaron-dominated carrier relaxation (see Figure 2b). Conversely, a recent investigation[17] has suggested that carrier relaxation is dominated by excitons. When an optical pulse creates electron and hole (*e-h*) pairs, strong Coulomb interactions lead to the formation of bound *e-h* pairs known as molecular excitons (Figure 2c). This study found that this charge-neutral exciton is the fundamental charge excitation that governs carrier relaxation in PANI (Figure 2d).

The primary aim of this paper is to determine the nature of photo-generated charge type in PANI. Our approach is to measure the time evolution of spin relaxation dynamics via ultrafast spin-resolved spectroscopy. When photo-excitation takes place in PANI, concurrent examination of the spin-dependent and spin-independent relaxation allows for determination of the dominant charge type. The two possibilities are clearly distinguishable by analyzing the spin-flip relaxation dynamics.[18-20] Specifically, as shown in Figure 2a,b, the polaron-dominated picture involves very fast spin relaxation decay due to an ultrafast spin flip process. Because the spin-polarized carriers are loosely bound (see inset of Figure 2a), the non-equilibrium spin-relaxation dynamics exhibit much faster carrier relaxation. In contrast, as displayed in Figure 2c,d, the corresponding spin-relaxation dynamics are substantially different for an exciton-dominated picture. In polymer physics, there exists a singlet and triplet state of exciton in PANI with different energy, and the corresponding spin conversion between the two states involves a spin flip process. Because complete spin conservation in exciton persists in a much longer time sale (a few microsecond)[21] than our measurement temporal window of a few picosecond, the spin conservation approximately represents a bounded nature (see inset of Figure 2c). No spin-flip process takes place during relaxation, so the spin-relaxation dynamics closely follow spin-independent relaxation. A comparison of spin-independent signals (via linearly-polarized pump and probe) and spin-dependent signals (via a combination of circularly-polarized pump and probe) clearly shows that the charge relaxation in conducting PANI exhibits no spin-flip process. This provides direct evidence of an exciton-dominated photo-response.

2. Experimental section

2.1. Preparation of PANI nanoparticles.

For the sample used in this experiment, water-stable PANI nanoparticles were prepared by solvent-shifting methods.[22] Briefly, 5 mg of PANI in EB state that fabricated by a chemical oxidation polymerization method, dissolved in 4 mL N-methyl-2-pyrrolidinone (NMP), later we added 30 mL aqueous solution containing 200 mg of Tween 80. The mixture was vigorously stirred at room temperature for 4 hours. After reacting, Tween 80-coated PANI nanoparticles were dialyzed to remove excess surfactants for 24 hours, and centrifuged using centricon (molecular weight cut off: 3,000 Da). The SEM image displayed in Figure 3a clearly shows the PANI nanoparticles. Colloidal size was 44.6 ± 10.6 nm, which was confirmed by dynamic laser scattering (see inset of Figure 3a). At concentration of $10^{-1}$ M HCl, the EB state PANI nanoparticles changed to the ES state. Absorbance spectra for the two groups were obtained by spectrometry (Figure 1b). The nanoparticle type of PANI minimizes a problem of conformational change, such as an aging issue.

2.2. Optical pump-probe spectroscopy.

The experimental schematic of our ultrafast pump-probe spectroscopy is displayed in Figure 3b. Ultrashort 60 fs, 3.1 eV pulses were delivered from 250 kHz Ti:sapphire regenerative amplifier (Coherent RegA 9050) via second-harmonic generation of 1.55 eV pulses through 1-mm-thick BBO crystal. Part of the amplifier output was used to generate probe pulses (white-light supercontinuum) on a 1-mm-thick sapphire disk. For the time-resolved pump-probe spectroscopy, the pump and probe pulses are focused on the sample in a non-collinear geometry in order to suppress the pump-induced scattered probe signals. The transmitted probe pulse is detected by a broadband near-infrared Si photodetector (Thorlabs PDA100A) after passing through a grating-equipped monochromer (Newport 74125) to perform spectrally-resolved measurements. The pump pulses are mechanically chopped before the sample at 10 kHz, and the differential transmission signals are measured using lock-in detection technique (Standard Research Systems SR850). The pump-probe delay is varied using a mechanical stepper stage (Newport M-IMS300PP). The two PANI samples were kept in a 3-mm quartz cell and actively stirred using a magnetic stirrer to prevent inter-particle aggregation effects occurs during measurements. All measurements were performed at room temperature.

3. Results and discussion

We first discuss measurements of spin-independent dynamics. Although this measurement generally does not involve spin-polarized charge dynamics, non-equilibrium characteristics are important for understanding carrier dynamics and developing polymer-based ultrafast optoelectronics.[14] EB and ES state PANI were excited by linearly polarized 3.1 eV pump pulses, and the non-equilibrium dynamics were measured (pump and probe were orthogonally polarized) (see the inset of Figure 3c). For ES state dynamics, the polaron population predominates due to proton doping. Because of this, the dynamics should be governed by polaron-carrier scattering exhibiting spin-independent relaxation features (spin-independent relaxation will be discussed later). Figure 3c directly compares the normalized differential transmission $\Delta T/T_0$ of EB (blue) and ES state PANI (green) as a function of the pump-probe delay $\Delta t$. The results clearly indicate that ES state PANI has a much faster relaxation component ($\tau_1$ = 0.24 ps) than EB state PANI ($\tau_1$ = 1.47 ps).

To understand these dynamics, two features should be addressed. First, the polaron-assisted relaxation mechanism is more dominant in ES state PANI than in EB state PANI. In contrast to the EB state, ES state PANI contains a larger number of polarons in the polymer chain before photo-excitation due to protonation doping.[5] The prevailed polaron population promotes the relaxation of the photo-generated charge carriers via collisions between equilibrium polarons and photo-generated carriers. Second, in ES state PANI, there is no potential barrier at the quinoid ring,[1] which naturally leads to enhanced charge transfer along the polymer chain. Excited carriers also move more freely in ES state than in EB state, and there is rapid restoration of the photo-bleaching effect. To transition from EB to ES state PANI, two important structural changes must occur: the creation of polarons and elimination of a potential barrier at the quinoid ring. Both of these changes contribute to efficient charge-transfer dynamics mediated by freely movable polarons along the chain.

Although the linearly polarized pump-probe measurements reveal non-equilibrium charge relaxation dynamics, they do not provide detailed information on the nature of *charge-excitation types*. In other words, analysis that relies on (spin-independent) time-resolved dynamics alone does *not* reveal the genuine character of photo-excited charge types. This is because slight changes in sample environments, such as sample type (film or solution) or differences during sample preparation, contribute to large fluctuations in time-domain signals due to the large dielectric mismatch. Previous studies have reported substantially different dynamics in time-resolved transient signals within a few picoseconds, and the results appeared to show inconsistent sample dependence.[16, 17]

To overcome this problem, ultrafast spin-resolved pump-probe spectroscopy was employed. The spin-resolved spectroscopy has been widely used in semiconductor physics to investigate spin relaxation dynamics of the excited charge carriers. These techniques provide insights into photo-generated charge types and carrier relaxation dynamics.[23-25] To determine the charge-excitation types, two sets of pump-probe measurements were performed: co-circularly polarized pump/probe measurements and anti-circularly polarized pump/probe measurements for ES state PANI (Figure 4a). We note that the circularly polarization of the beam in one direction (clockwise or counter-clockwise) results in preferential optical excitation with only one type of spin component.

Figures 4b and 4c show spectrally resolved $\Delta T/T_0$ for different $\Delta t$ with co-circularly and anti-circularly polarized pulses. Note that the co-circularly polarized pump-probe signals (Figure 4b) exhibit substantial spectral peaks at a center photon energy of 1.59 eV, whereas peaks are not observed in the anti-circularly polarized pump-probe signals (Figure 4c). From the simplest point of view, a differential probe spectrum arises from changes in spin-polarized carrier occupation. After carrier excitation with one type of spin, the corresponding spin-polarized occupation probability increased and absorption of the co-circularly polarized probe was reduced. This is a classic hallmark of the spin Pauli-blocking effect, which provides direct evidence that charge-excitation is strongly spin-polarized as schematically explained in Figure 4d. It is interesting to note that the co-circularly polarized pump-probe spectrum (Figure 4b) is similar to the absorption spectrum (see Figure 1b). This feature would be different if the photo-excited charges were not spin-polarized.

More insights were obtained by inspecting the sum and subtraction of the spin-resolved signals. Three main aspects are addressed here. First, the pure spin-relaxation dynamics of photo-generated carriers are obtained by subtracting anti-circularly polarized signals from co-circularly polarized signals ($S_-$). The sum of the two signals ($S_+$) represents spin-independent relaxation dynamics. The inset of Figure 4b, c directly compares the two $S_\pm$ signals as a function of pump-probe delay. Note that the spin-dependent relaxation dynamics ($S_-$) closely follows the spin-independent ($S_+$) signals. Because the $S_+$ signals do not involve spin imbalance in photo-excited carriers, the corresponding data do not exhibit spin-dependent relaxation. The $S_-$ dynamics exhibited photo-response restoration dynamics from a spin-imbalanced system to a (spin-balanced) equilibrium system.[23] Here, the $S_-$ dynamics involve two different spin-relaxation pathways: the usual carrier recombination process to the ground state and the spin-flip process, in which total carrier densities are preserved in the excited state.[25] If the spin-flip process is dominant in the $S_-$ dynamics, they should accompany a faster relaxation component than the $S_+$ dynamics. As shown in the inset of Figure 4b, c, $S_+$ and $S_-$ are nearly indistinguishable, illustrating that there is no spin-flip process involved in the relaxation dynamics. It has been well established that for spin-polarized bound particles such as excitons, no spin-flip process occurs during relaxation. In contrast, loosely bound polarons experience efficient spin-flip processes, as evident in previous works.[18-20, 26] As a result, the non-equilibrium relaxation dynamics are dominated by excitons. Second, as shown in Figure 4e, we note that the overall magnitude of linearly polarized pump-probe signals at 1.59 eV is approximately same as the sum of the spectra with co-circularly and anti-circularly polarized pump and probe pulses (see Figure 4b,c), which is consistent with a relationship between linearly and circularly polarized set, namely the linear polarization can be decomposed into co-circular and anti-circular polarization. Finally, for the time-resolved dynamics (Figure 4f), the linearly polarized pump-probe signals closely follow the spin-independent dynamics ($S_+$), i.e. the $S_+$ signals show the same dynamic behavior as the spin-independent signals (green in Figure 3). This is expected because the linearly polarized pump-probe signals do not contain any spin imbalance information in the photo-excited carriers.

4. Conclusion

To conclude, we investigated non-equilibrium charge dynamics in PANI. Spin-independent spectroscopy demonstrated that the photo-response in ES state PANI exhibits faster relaxation than EB state PANI. This arises from scattering between the predominant polaron and photo-excited carriers. Ultrafast spin-resolved pump-probe spectroscopy revealed that charge relaxation is dominated by excitons. By investigating the spin-dependent and spin-independent dynamics, we provided experimental evidence that the spin Pauli-blocking effect, relaxation time, and the absence of a spin-flip process underlie the exciton-dominated charge types in PANI.

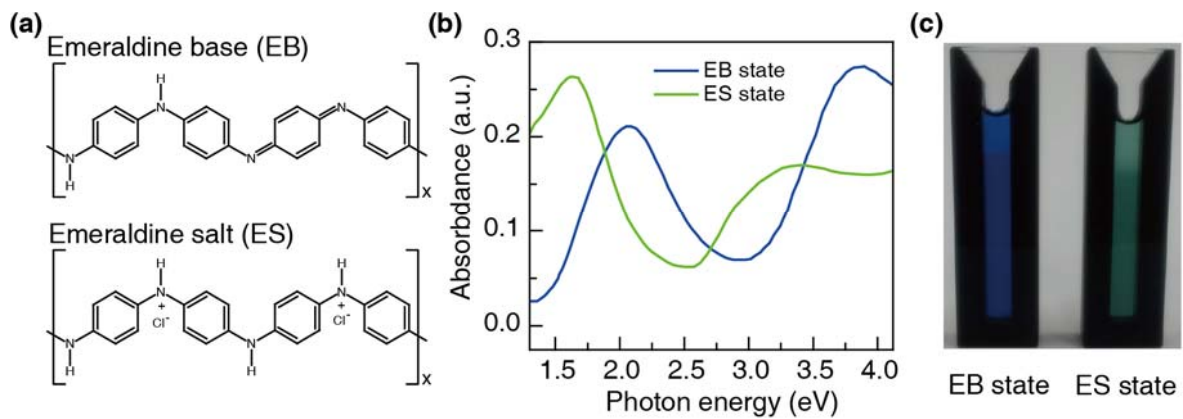

**Figure 1**. (a) Chemical structures of PANI nanoparticles for EB and ES states. The phase transition from EB state to ES state is governed by proton-doping. For the transition to ES state of PANI nanoparticles, proton doping supplies positive charge carriers along the polymer chain. (b) The optical absorption spectra of EB and ES state PANI nanoparticles. A strong red-shifted feature due to π-π* transition of benzenoid rings is clearly visible. (c) Picture of EB and ES states for PANI nanoparticles.

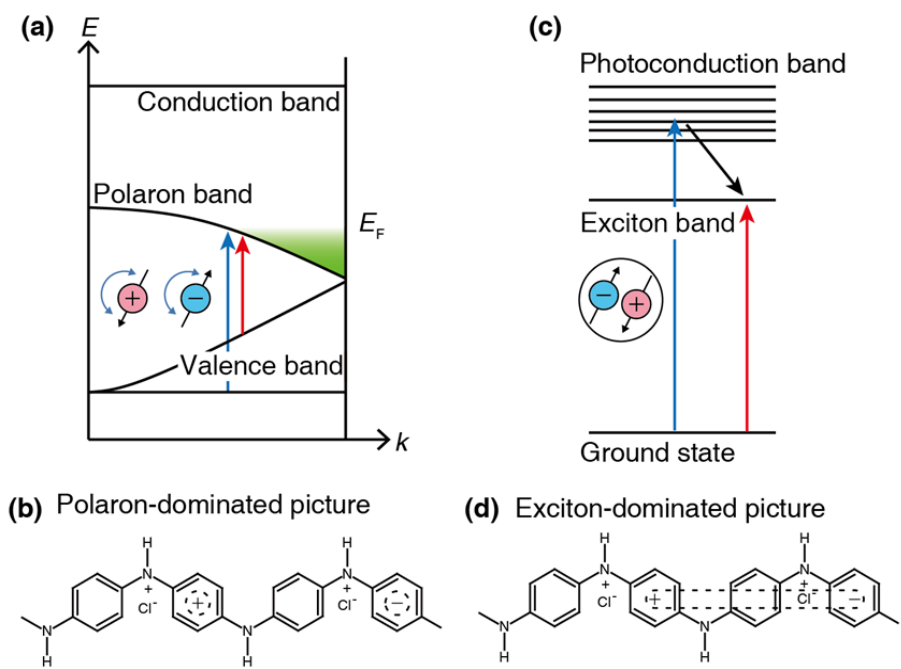

**Figure 2**. (a) Schematics representing photo-induced non-equilibrium dynamics with polaron-dominated charge relaxation in ES state PANI. For the band-structure concept, the pump beam (blue arrow) and probe beam (red arrow) induce optical transition from valence bands to the polaron band. The inset shows schematic for representing loosely bounded spin nature of polarons. (b) The structural form of the polaron-dominated picture in ES state PANI. (c) In the exciton-dominated picture, photo-excitation provides *e-h* pairs, which rapidly relax and form exciton by Coulomb interaction. In contrast to the polaron-dominated picture, the exciton-dominated picture represents strongly bound spin-polarized charge relaxation. The inset shows schematic for the strongly bounded exciton (d) The structural form of the exciton-dominated picture in ES state PANI.

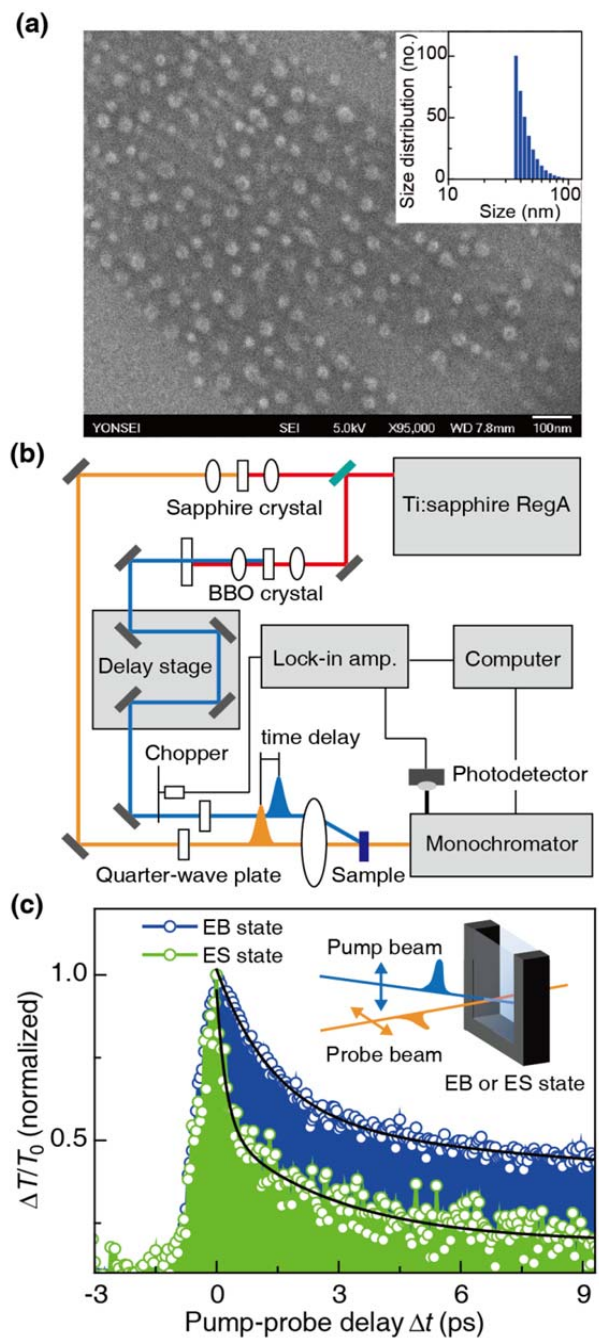

**Figure 3**. (a) SEM image of the PANI nanoparticles. The inset shows size distribution of the PANI nanoparticles. (b) Experimental scheme of pump-probe spectroscopy. (c) Normalized differential transmission $\Delta T/T_0$ dynamics measured by linearly polarized pump and probe spectroscopy are displayed as a function of pump-probe delay $\Delta t$ for EB (blue) and ES (green) state PANI nanoparticles. ES state PANI nanoparticles exhibit a rapid decay component that is not visible in EB state dynamics. Black lines show bi-exponential fits. The inset shows

schematics of pump-probe experiment. The polarization of the pump and probe pulse is orthogonal (linearly polarized).

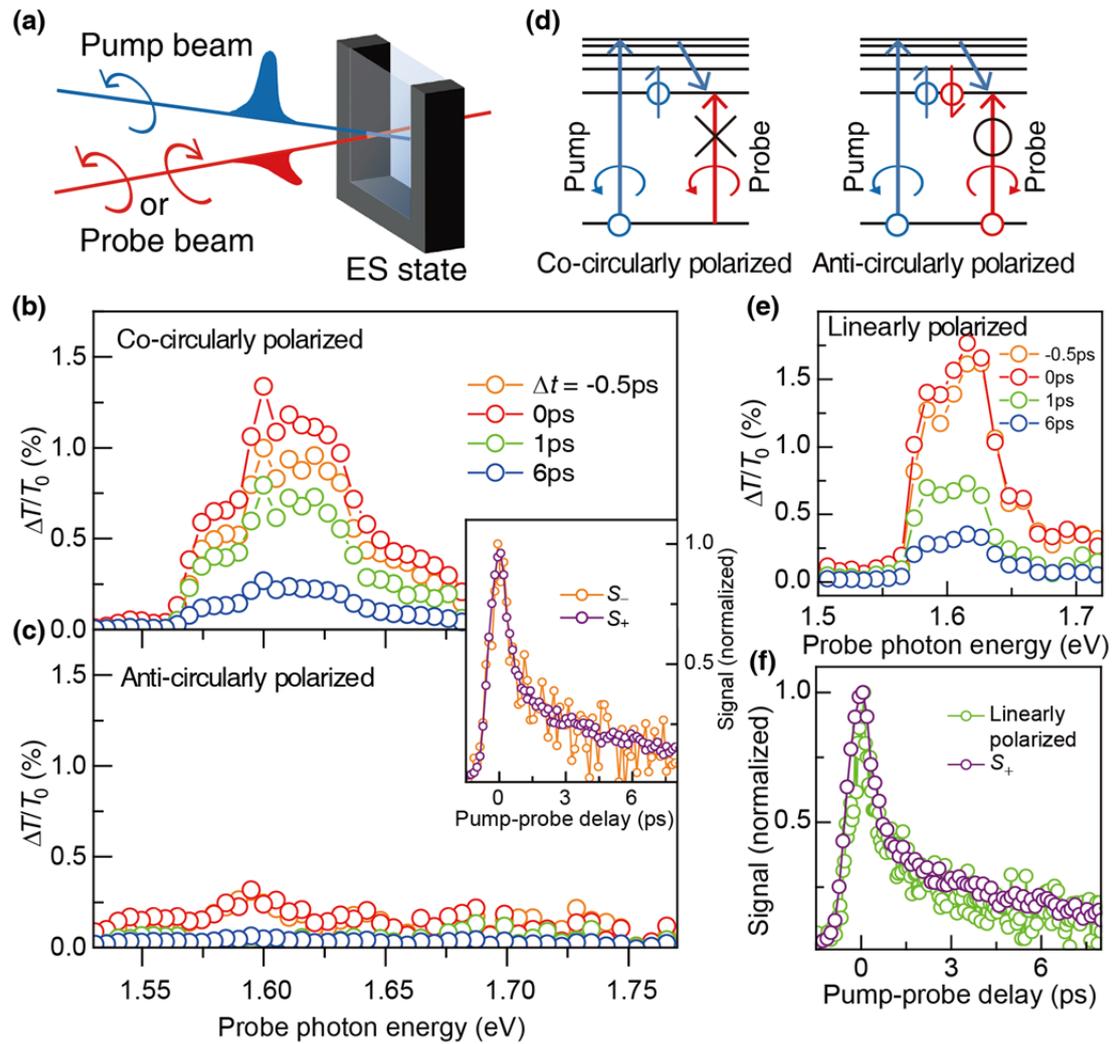

**Figure 4**. (a) Schematics of spin-resolved pump-probe spectroscopy with circularly-polarized pulses for ES state PANI. (b, c) Spectrally-resolved $\Delta T/T_0$ for several $\Delta t$ with (b) co-circularly and (c) anti-circularly polarized pump-probe pulses. The inset shows normalized $S_\pm$ signals as a function of $\Delta t$. The probe photon energy is 1.59 eV. (d) Schematics representing Pauli-block effect with co-circularly polarized pump-probe sets. No spin Pauli-blocking is expected for anti-circularly polarized pump-probe spectroscopy. (e) Normalized spectrally-resolved $\Delta T/T_0$ for several $\Delta t$ with linearly polarized pump-probe pulses. (f) Comparison of the $S_+$ and the linearly polarized pump-probe signals for the ES state PANI is displayed.

AUTHOR INFORMATION

**Corresponding Author**

*E-mail: YMHUH@yuhs.ac (Y-M.H) and hychoi@yonsei.ac.kr (H.C).

**Author Contributions**

The manuscript was written through contributions of all authors. All authors have approved the final version of the manuscript.

**Funding Sources**

The authors declare no competing financial interests.

ACKNOWLEDGMENT

The work at Yonsei was supported by the Basic Research Program through the National Research Foundation of Korea (NRF), funded by the Ministry of Education, Science, and Technology (No. 2011-0013255), the NRF grant funded by the Korean government (MEST) (NRF-2011-220-D00052, No. 2011-0028594, No. 2011-0032019, No. 2010-0023202), and the LG Display Academic Industrial Cooperation Program.